\documentclass[10pt,letterpaper]{article}
\usepackage[top=0.85in,left=2.75in,footskip=0.75in,marginparwidth=2in]{geometry}

\usepackage[utf8]{inputenc}

\usepackage{cite}

\usepackage{nameref,hyperref}

\usepackage[right]{lineno}

\usepackage{microtype}
\DisableLigatures[f]{encoding = *, family = * }

\raggedright
\usepackage{changepage}

\usepackage[aboveskip=1pt,labelfont=bf,labelsep=period,singlelinecheck=off]{caption}

\makeatletter
\renewcommand{\@biblabel}[1]{\quad#1.}
\makeatother

\usepackage{lastpage,fancyhdr,graphicx}
\usepackage{epstopdf}
\fancyheadoffset[L]{2.25in}
\fancyfootoffset[L]{2.25in}

\usepackage{color}

\definecolor{Gray}{gray}{.25}

\usepackage{graphicx}

\usepackage{sidecap}

\usepackage{wrapfig}
\usepackage[pscoord]{eso-pic}
\usepackage[fulladjust]{marginnote}
\reversemarginpar

\usepackage{enumitem}

\usepackage{tikz}
\usetikzlibrary{shapes.geometric,arrows, decorations.markings, decorations.pathreplacing, shapes.misc,positioning,calc}
\tikzset{cross/.style={cross out, draw=black, minimum size=2*(#1-\pgflinewidth), inner sep=0pt, outer sep=0pt},cross/.default={0.4cm}}
\usepackage{adjustbox}
\usepackage{svg}

\begin{document}
\vspace*{0.35in}

\begin{flushleft}
{\LARGE
\textbf\newline{Graphical models for inference: A model comparison approach for analyzing bacterial conjugation}
}
\newline
\\
Nat Kendal-Freedman\textsuperscript{1\dag\textcurrency*},
Joseph Victor Fiorillo Meleshko\textsuperscript{2\dag*},
Aaron Yip\textsuperscript{3},
Brian Ingalls\textsuperscript{1}
\\
\bigskip
\textbf{1} Department of Applied Mathematics, University of Waterloo, Waterloo, Ontario, Canada \\
\textbf{2} David R. Cheriton School of Computer Science, University of Waterloo, Waterloo, Ontario, Canada\\
\textbf{3} Department of Chemical Engineering, University of Waterloo, Waterloo, Ontario, Canada 
\\
\bigskip
\dag These authors contributed equally to this work.\\
\textcurrency Current Address: Max Planck Institute for Mathematics in the Sciences, Leipzig, Saxony, Germany\\
\textasteriskcentered Corresponding Authors: nmkendal@uwaterloo.ca, jmeleshk@uwaterloo.ca

\end{flushleft}

\section*{Abstract}
We present a proof-of-concept of a model comparison approach for analyzing spatio-temporal observations of interacting populations. Our model variants are a collection of structurally similar Bayesian networks. Their distinct Noisy-Or conditional probability distributions describe interactions within the population, with each distribution corresponding to a specific mechanism of interaction. To determine which distributions most accurately represent the underlying mechanisms, we examine the accuracy of each Bayesian network with respect to observational data. We implement such a system for observations of bacterial populations engaged in conjugation, a type of horizontal gene transfer that allows microbes to share genetic material with nearby cells through physical contact. Evaluating cell-specific factors that affect conjugation is generally difficult because of the stochastic nature of the process. Our approach provides a new method for gaining insight into this process. We compare eight model variations for each of three experimental trials and rank them using two different metrics.

\section{Introduction}

Microbes transfer genetic information to other cells through several mechanisms, collectively referred to as horizontal gene transfer (HGT) (distinct from vertical gene transfer to progeny). Conjugation is a HGT process in which bacteria share genetic information through a physical connection. It involves the transfer of a replicable, extra-chromosomal DNA molecule called a plasmid~\cite{davison99,norman09,virolle20}. Plasmids often carry  accessory genes that benefit their hosts in specific environmental conditions. (A well-known example is antibiotic resistance plasmids.) Because the plasmid replicates in the host and can be passed on, the spread of a plasmid through a bacterial population is analogous to the spread of a virus through a human population.

Conjugation plays an important role in prokaryotic evolution~\cite{davison99} and in the development and maintenance of biofilms~\cite{madsen12, virolle20}. The capacity to design and construct self-propagating plasmids makes conjugation a promising tool for genetic engineering of microbial communities~\cite{padilla15,yip22}, with applications to health, environmental remediation, wastewater treatment, and agriculture. However, the risks inherent in environmental release of modified, mobilizable genetic elements require careful consideration~\cite{padilla15}. It would be beneficial to gain insight into the mechanisms governing conjugation at the single-cell level and how they impact spread of a plasmid through a population.

\begin{figure}[]
    \centering
    \includegraphics[width=0.7\textwidth]{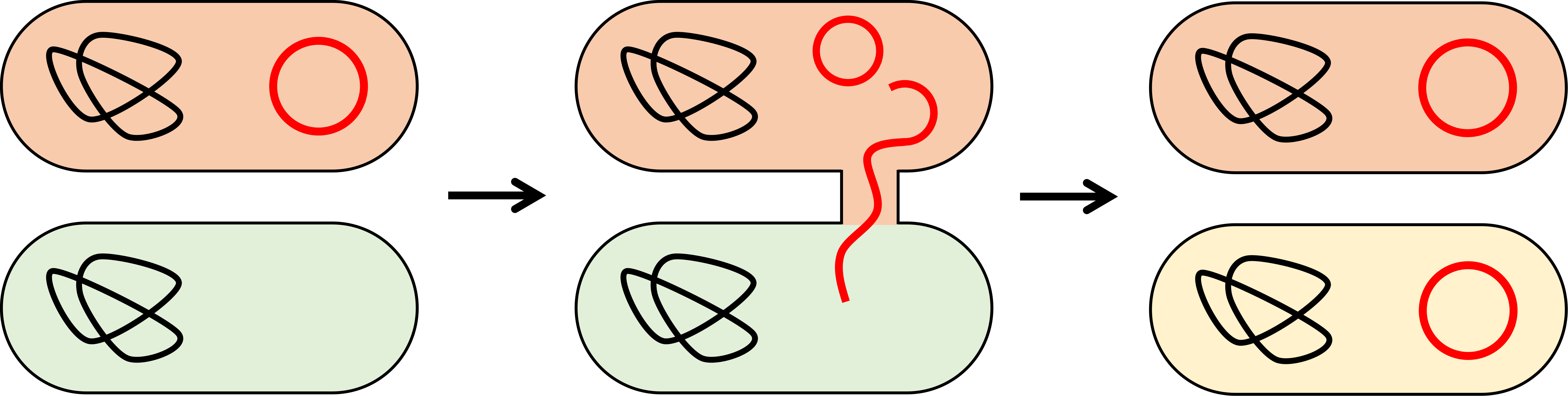}
    \caption[Diagram of Conjugation]{A donor cell (orange) transfers a plasmid (orange DNA; black is chromosome) to a recipient cell (green), causing it to change phenotype (transconjugant; green).}
    \label{fig:conjugation}
\end{figure}

The process of conjugation can be summarized as follows~\cite{norman09,davison99} (Fig~\ref{fig:conjugation}). A donor cell containing the plasmid extends a tube-like appendage called a pilus. The pilus facilitates a physical connection with a recipient cell, allowing a copy of the plasmid to be transferred to the recipient. The connection is then severed, and the plasmid replicates and initiates production of conjugative machinery in the recipient. A recipient that has received a plasmid is called a transconjugant. While the conjugation event can occur on a timescale of minutes, there may be substantially longer delays before a transconjugant expression the functional and conjugative genes on the plasmid~\cite{massoudieh07}. These delays pose interesting interpretation challenges which we aim to address.

\subsection{Challenges}

Bacterial conjugation has been researched extensively. As it is difficult to study and model at the single-cell level, many of these investigations have focused on population-level assays~\cite{sorensen05,yip22}. Reporter genes (such as fluorescent proteins) can be used to track the spread of a plasmid through a population. However, rapid population growth makes it challenging to distinguish whether gene spread is due to conjugation or to cell division. Population-level traits such as conjugation frequency, ratios of donor to recipient cells, and environmental conditions~\cite{johnsen07} are frequently reported. The effects of various factors on the spatial organization of donor, recipient, and transconjugant cells have also been investigated~\cite{ruan24,ma23,ma24}. Recent advances in microfluidics led to the development of traps in which populations of cells grow in a single layer~\cite{luan20,yin12}. Time-lapse imaging of these traps provides information on individual cell features, which can be used to study how a plasmid spreads through individuals in a population spatio-temporally. Even with these advances, the scale at which conjugation occurs and the variable delays during the process make it challenging to identify and analyze individual conjugation events~\cite{massoudieh07,seoane11}.

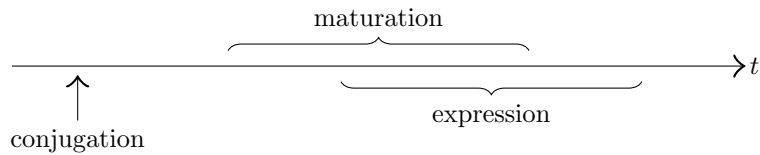
\begin{figure}[]
    \centering
    \begin{tikzpicture}
        \node[] (start) at (0, 0) {};
        \node[] (end) at (10, 0) {};
        \node[] at (10,0) {$t$};
        \node[] (label) at (1,-1) {conjugation};
        \node[] (conj) at (1,0) {};

        \draw [decorate,decoration={brace,amplitude=5pt,mirror,raise=4ex}]
                (7,-0.4) -- (3,-0.4) node[midway,yshift=3em]{maturation};

        \draw [decorate,decoration={brace,amplitude=5pt,mirror,raise=4ex}]
                (4.5,0.4) -- (8.5,0.4) node[midway,yshift=-3em]{expression};
        
        \foreach \from/\to in {start/end, label/conj}
        \draw[decoration={markings,mark=at position 1 with
        {\arrow[scale=2.5,>=to]{>}}},postaction={decorate}] 
        (\from) -- (\to);
    \end{tikzpicture}
    \caption[Delays in the Conjugative Process]{A potential timeline of delays in the conjugative process. Brackets denote potential time ranges in which an expression/maturation threshold is reached.}
    \label{fig:delays}
\end{figure}

The most relevant delays are the `expression' and `maturation' delays that occur after a conjugation event (Fig~\ref{fig:delays}). We define the expression delay as the length of time between the conjugation event (presumed instantaneous) and the time at which expression of the reporter gene is observed. Likewise, the maturation delay is the length of time between the conjugation event and the time at which the transconjugant begins acting as a donor (i.e.~matures). These unknown delays make it challenging to determine both the timing of events and the set of potential donors. For instance, a cell may donate the plasmid before it is identified as a transconjugant (analogous to an infected person spreading an infection before they are symptomatic).

Several studies have relied on ad hoc, manual identification of donor-recipient pairs~\cite{babic08,lawley02,li19,liu23,seoane11}. These studies have restricted themselves to cases for which there is only one potential donor for a particular transconjugant. They have provided valuable insight into how factors such as the distance between cells and the relative orientation of cells affect conjugation. More recent studies use fluorescent proteins that bind to single or double stranded DNA to detect exact donor-recipient pairs and investigate other properties surrounding the intracellular dynamics of conjugation ~\cite{couturier23,goldlust23}.

\subsection{Modelling Approaches}

Manual identification of conjugation events is both time-consuming and limited to clear-cut situations. Mathematical models provide a strategy to complement these data sets with limited information. Both deterministic and stochastic approaches have been used to model conjugation and interpret experimental data at the population level. They have been used to investigate the delays in the conjugative process~\cite{massoudieh07}, plasmid persistence~\cite{ponciano07}, and colony interactions~\cite{lagido03}. At the single-cell level, several agent based models (ABMs) of conjugation have been developed~\cite{cellmodeller, discus, cosmic}. ABMs are suitable for modelling processes with large, stochastic, and heterogeneous populations, but require extensive knowledge of the underlying processes and are computationally expensive. Notably, the models published to date do not vary the probability of conjugation based on properties of the donor-recipient pair. 

Here, we present probabilistic graphical models (PGMs) as a novel approach to modelling conjugation. PGMs have been used to model other biological systems, including gene networks and protein expression~\cite{ni18}. A PGM is a graphical representation of a joint distribution of a set of random variables that can facilitate efficient queries of the distribution. (A query is the computation of the likelihood of some assignment to a subset of the random variables, conditioned on an assignment to a disjoint subset of the random variables given as evidence.) Examining a PGM from a graph-theoretic perspective simplifies the probability calculation by leveraging independence relations between the random variables. Understanding the potential set of interactions is necessary to construct a PGM, after which various techniques can be used to investigate the parameters governing their likelihoods.

A Bayesian network (BN) is a type of PGM in which the representation is a directed, acyclic graph. BNs are particularly useful for modelling causal effects. They have been applied to several problems within mathematical biology, particularly in the context of machine learning~\cite{needham07, ni22}. In applications to healthcare, these models have been used to analyze data, estimate parameters, and explore potential interactions~\cite{kyrimi21}. Although BNs are not currently being used in clinical settings, they have the potential to be a powerful tool for clinical decision making for diagnosis and treatment. Another potential area of application is social networks, such as social media. BNs can be used to infer information about individuals, recommend products to users, and study the flow of information through a network~\cite{farasat15}. 

\subsection{Bayesian Networks}

A PGM consists of two components: a graph structure and a compatible set of conditional probability distributions (CPDs). The graph structure represents a set of interdependent random variables by identifying each variable with a node and CPD and representing each dependency between variables with an edge. The CPD of a random variable in a BN is a distribution for that variable, and it is conditioned on its parents in the graph. The set of all such CPDs must factorize the graph; that is, the product of the entire set of distributions must equal the joint distribution of the random variables. Factorization allows for piecewise calculation of a query of the joint distribution and reveals the independence relationships among the random variables. In this paper, we use \textit{query} to mean a probability query $P(Y | E = e)$ over a set of random variables $X$, where $Y, E \subseteq X$, $Y \cap E = \emptyset$, and $e$ is an assignment of $E$. Further detail on factorization and independence relationships in this context can be found in~\cite[Ch.~2]{koller09}.

\begin{figure}[]
    \centering
    \begin{tikzpicture}[state/.style={circle, draw, minimum size=1cm}]
        \node[state] (X1) at (0.5, 2) {$v_1$};
        \node[] at (1.5, 0.75) {$p_{v_1}$};
        \node[state] (X2) at (3, 3) {$v_2$};
        \node[] at (3.4, 1.5) {$p_{v_2}$};
        \node[state] (X3) at (5.5, 2) {$v_3$};
        \node[] at (4.5, 0.75) {$p_{v_3}$};

        \node[state] (Y) at (3, 0) {$x$};
        
        \foreach \from/\to in {X1/Y, X2/Y, X3/Y}
        \draw[decoration={markings,mark=at position 1 with
        {\arrow[scale=3,>=to]{>}}},postaction={decorate}] 
        (\from) -- (\to);       
    \end{tikzpicture}
    \caption[Depiction of a Noisy OR Statement]{A graph representation of a noisy OR statement. The random variable $x$ has a non-zero probability of being true if at least one $v_i=1$ with associated $p_{v_i} > 0$. The edge weights are are given by $p_{v_i} = P(x=1|v_i=1)$.}
    \label{fig:noisy_or}
\end{figure}
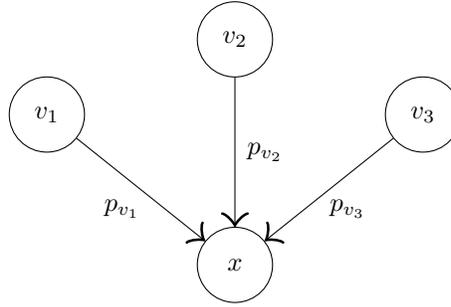

Calculation of queries in a BN can be computationally intractable, even given an optimal factorization. One solution is to constrain the BNs to binary random variables, which limits the number of possible assignments. Additionally, it is possible to use a \textit{Noisy-OR} formulation for the CPDs. A (binary) Noisy-OR CPD describes the distribution for a random variable $x$ as a logical OR of its parents, up to some \textit{noise}. In this context, the variables on which $x$ depends can be viewed as causes; if any of them have value 1, then $x$ will also have value 1 unless the noise inhibits the implication. The Noisy-OR CPD has a probability $p_v$ associated to each parent variable $v$ (Fig~\ref{fig:noisy_or}). Even if a parent variable $v$ has value 1, it only causes $x$ to be 1 with probability $p_v$. If the parent variable has value 0, then it will never cause $x$ to be 1 regardless of noise. This relationship can be computed and stored far more efficiently than a general CPD with arbitrary values. Furthermore, using exclusively Noisy-OR CPDs simplifies the computations involved in taking the product of CPDs and leads to several heuristic speed improvements.

While limiting ourselves to Noisy-OR CPDs may seem overly restrictive, they can meaningfully represent many systems. For example, in health, symptoms can be modelled as a Noisy-OR of the diseases that may cause them. In our case, the (binary) state for a cell is having the plasmid at a particular time, which can be modelled as a Noisy-OR of the neighbouring cells that could conjugate to it along with the state of having already received it in the past.

Here, we present a proof-of-concept of a novel approach for inferring information about cell-cell interactions -- conjugation events -- from spatio-temporal data. Section~\ref{sec:methods} includes a brief overview of the experimental setup as well as the construction of our models. We frame the information gathered from time-lapse data as a BN with Noisy-OR CPDs. Relevant cell features are represented as nodes, and causal effects from contacts or lineages are represented by edges. We then cover how we evaluate and compare the model versions in Section~\ref{sec:comparison}. Model variants are created by changing the probability distributions that govern the interactions. Tested functions and results are included in Section~\ref{sec:results} and a discussion of them is included in Section~\ref{sec:discussion}. We demonstrate how our method can be used to gain insight into the biological mechanisms governing conjugation. 

\section{Methods \& Models} \label{sec:methods}

\subsection{Experimental Design}

\begin{figure}[]
    \centering
    \includegraphics[width=0.5\textwidth]{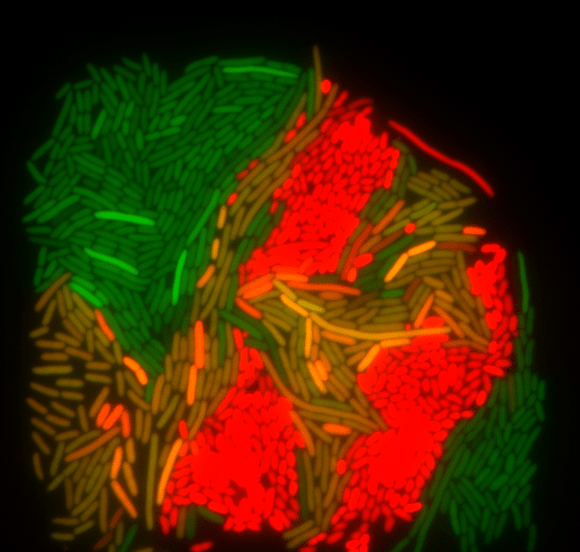}
    \caption[Sample Frame from an Experiment]{A sample frame in which recipients (green), donors (red), and transconjugants (orange) are visible. Relative fluorescence levels have been adjusted for visibility.}
    \label{fig:frame}
\end{figure}

Data for this work consists of currently unpublished time-lapse microscopy images. Briefly: two populations of~\textit{Escherichia coli} cells are grown in a single layer inside microfluidic traps. Donor cells contain a conjugative plasmid that codes for a red fluorescent protein (RFP). Recipient cells contain a non-conjugative plasmid that codes for a green fluorescent protein (GFP). As the populations interact, RFP-carrying plasmids are transferred to the recipient population, leading to the formation of transconjugant cells which appear orange (carrying both color signatures) after an expression delay. Images are taken in 5 minute intervals over a span of 20-24 hours. A representative frame is shown in Fig~\ref{fig:frame}. A combination of preexisting software~\cite{cellprofiler,omnipose} and custom software developed is used to extract relevant information from the images~\cite{labgithub}. A more detailed description of the image processing pipeline can be found in~\cite{ahmadi24}.

After pre-processing, fluorescence values are used to determine cell type. Ilastik~\cite{ilastik}, a machine learning tool for image classification, was used to identify newly formed transconjugant cells. All descendants of any cell labelled as a transconjugant are also considered to be transconjugants.

\subsection{Assumptions}

The experimental data allows us to investigate the impact of spatial positioning and individual cell features on conjugation frequency, as well as the delays in the conjugative process. We make the following biological assumptions based on the setup of the experiment:

\begin{enumerate}[label=(\roman*)]
    \item \textbf{No Loss of Properties:} If a cell has the plasmid, is mature, or expresses RFP, then all of its descendants have that property. Similarly, the cell will not lose the property for the remainder of its existence.
    \item \textbf{Instantaneous Conjugation:} Conjugation has been observed to occur on the same time scale as images are taken (5 minutes)~\cite{andrup98,andrup99}. Thus, we assume conjugation occurs `instantaneously' within a frame.
    \item \textbf{Adjacency during Conjugation:} The experiment used a P-plasmid system which is characterized by short, rigid pili and relies on direct cell wall contact~\cite{jass04,samuels00}. Because cells are tightly packed during the majority of the experiment, we assume that donor-recipient pairs must be in contact.
    \item \textbf{Initial Conditions:} All cells expressing RFP in the first time frame are mature (i.e.~can donate), and all cells not expressing RFP do not contain the plasmid (i.e.~are recipients). 
\end{enumerate}

The following assumptions are necessary for the model formulation and evaluation.

\begin{enumerate}[resume,label=(\roman*)]
    \item \textbf{Cell Properties:} The same governing functions are used for every cell of the same type. In particular, we do not assume individual cells are inherently more likely to donate/receive plasmids \cite{andrup98,andrup99,cullum78}. 
    \item \textbf{Potential Recipients:} Conjugation can only target recipient cells. (While secondary conjugation events are possible, we cannot detect them from the available data.) 
    \item \textbf{Delay Ranges:} After receiving a plasmid through conjugation, a transconjugant lineage undergoes an expression delay before an RFP signal appears, and a maturation delay before it can donate a copy of the plasmid. These delays are assumed to be at least five minutes (the frame-to-frame time interval). This assumption ensures that the BN graph structure is acyclic. 
\end{enumerate}

Plasmid loss is considered separately, as accounting for loss is occasionally necessary to resolve issues. If a cell divides shortly after being conjugated to, it may contain few copies of the plasmid and pass them on to a single daughter lineage. This situation can lead the model may derive a contradiction due to Assumptions (i) and (vii). In such a case, we ignore Assumption (i) for the lineage which did not light up and assume that it has lost the plasmid. Doing so effectively disconnects the cell from the parent lineage that must have received the plasmid.

\subsection{Graph Structure}

In the context of the conjugation experiment we examine, the nodes in the Bayesian network (BNs) represent random variables corresponding to properties of cells. Specifically, these are properties that will either be inferred by the model or provided as evidence. The expression of fluorescent proteins is used to infer whether cells carry the plasmid and whether they are mature. Each cell in each image frame is represented by a unique triple of binary random variables $(g,m,r)$ that represent whether the cell has the plasmid (gene), is mature, and is displaying RFP intensity above a threshold, respectively. The conditional probability distribution (CPD) of each of these binary random variables is given as a Noisy-OR. Other cell features which may impact conjugation frequency are accounted for when determining the noise probability for the Noisy-OR CPDs. The noise probabilities can be thought of as \textit{edge weights} in the BN. 

The benefit of using a probabilistic graphical model (PGM) is that it implicitly accounts for how one conjugation event impacts the probability of other potential events. For instance, the probability that a cell receives the plasmid and matures impacts the probability that it conjugates to other cells in the future. Conversely, a cell is slightly less likely to have the plasmid or be mature if it is in contact with many recipients that do not begin expressing RFP afterwards. We discuss our specific construction that leverages these effects in the remainder of this section.

The edges in a BN represent potential interactions between variables, i.e.~cell properties. When discussing interactions within a lineage, we use the term `descendant' to refer to the corresponding cell in future time frames as well as its children. We use the term `cell' to refer to a cell in a single image frame. The relevant biological interactions can be summarized as follows:

\begin{enumerate}[label=(\roman*)]
    \item The properties of a cell impact the corresponding properties of its descendant(s) in the next frame.
    \item Contacts impact the probability a cell gains the plasmid through conjugation. The set of cells considered to be in contact can vary between models, see Section~\ref{sec:functions}.
    \item Receiving a plasmid impacts the probability that a cell's descendants express RFP and become mature. 
\end{enumerate}

To model these effects, we include the following types of edges. To simplify the construction of the graph, edges are determined without referencing the properties of each cell. (The graph will include unnecessary edges.) 
\begin{itemize}
    \item \textbf{Type (i)}: There is an edge pointing from each cell's gene, maturation, and RFP nodes to the corresponding nodes of its direct descendant(s). 
    \item \textbf{Type (ii)}: Whenever two cells are in contact, there is an edge pointing from each cell's maturation node into the other cell's gene node.
    \item \textbf{Type (iii)}: There are edges pointing from each cell's gene node to the maturation and RFP nodes of its descendants within the relevant time ranges.
\end{itemize}

Type (iii) edges account for the maturation and expression delays. Because the delay is not fixed, edges point into every descendant within a specified delay range. In principle, all edges that are valid in some theoretical model should be included in the BN. Doing so would require the addition of an extremely large number of irrelevant edges. For instance, we would include a type (iii) edge for maturation from a cell in the first frame to all of its descendants in the entire experiment. In a realistic model, the maturation delay is bounded and most of these edges have zero weight. For computability and interpretability, we omit any edge with weight zero. One consequence of this choice is that the graph construction is dependent on the choice of the feasible ranges. Note also that the construction is guaranteed to be acyclic: both type (i) and type (iii) edges point forward in time and therefore cannot form a cycle. Type (ii) edges are within the same time step, but cannot form a cycle because they only point from maturation nodes into gene nodes. 

\subsection{CPDs \& Edge Weights}

The types of edges also highlight the ways in which we can vary the CPDs used to generate our model. Cells that have a property always pass it on to their descendants, so type (i) edges have weight 1. (They are certain to cause their descendant to have a specific property.) However, we consider different functions governing the likelihood of conjugation and different probability distributions for the maturation and expression delays. 

The conjugation probability functions can account for the spatial positioning of the pair as well as individual cell features. Because the baseline probability of conjugation is unknown, this function does not attempt to model the true biological probability of conjugation. Rather, it will be a relative measure of likelihood - pairs with higher values are considered more likely to conjugate. This relative measure is enforced by a normalization over all non-trivial type (ii) edges in model. A type (ii) edge is non-trivial if it represents an edge into a cell not currently expressing RFP. Because the experimental setup does not provide evidence of secondary conjugation events, we cannot assess the likelihood a trivial edge resulted in conjugation. To avoid penalizing distributions for recognizing good contacts which cannot lead to an observable conjugation event, we exclude trivial edges from the normalization.

The delays can be modelled by cumulative distribution functions (CDFs) that represent how likely the lineage is to have matured or begun expressing RFP as a function of the time since conjugation. The values given by the CDFs are then embedded into the Noisy-OR CPDs of the maturation and RFP nodes. Therefore, there are three main functions which determine a model: the conjugation probability function, the maturation delay function, and the expression delay function.

In summary, the weights for each type of edge are determined as follows:
\begin{enumerate}[label=(\roman*)]
    \item If a cell has a given property ($g$, $r$, or $c$ $=1$), then its descendants have the same property. Therefore, all edges from a cell to its direct descendant(s) have weight one.
    \item The probability of conjugation occurring between a pair of cells is independent of any other edges. The weight of each edge is the (relative) likelihood of conjugation.
    \item The probability that a cell reaches maturity (or reaches the fluorescence threshold) is based on a cumulative distribution. Embedding it into a Noisy-OR CPD requires a more complicated calculation, as described below.
\end{enumerate}

Maturation and expression edge weights are calculated inductively. It is necessary to account for the probability that none of the earlier edges pointing out of the same gene node caused the change. Let $f(t)$ be the CDF for a delay, $\alpha_t$ be the weight of the edge corresponding to a delay of $t$ minutes, and ${t_1,\ldots,t_n}$ be the times of the image frames within the potential delay range. Note that $\alpha_{t_1} = f(t_1)$ because there are no previous possibilities of maturation/expression to account for. The probability that no edge before time $t_k$ led to a change is given by $\prod_i=1^{k} (1 - \alpha{t_i})$ and must equal the value specified by the CDF ($1-f(t_k)$). Rearranging the equality 
$$1 - f(t_k) = \prod_{i=1}^{k}(1-\alpha_i)$$
gives us the formula
$$\alpha_k = 1 - \frac{1 - f(t_k)}{\prod_{i=1}^{k-1}(1-\alpha_i)}.$$

\section{Model Evaluation \& Comparison} \label{sec:comparison}

Our approach is designed to allow inference from experimental data. To do so, we create a set of models using the above approach. Each model is based on a different combination of parameters and functions describing the conjugative process. We can then compare models to see which one `best captures' the experimental data. For a single observed conjugation event (i.e.~appearance of above-threshold RFP fluorescence), the best model is the one that assigns the highest probability to that event when conditioned on experimental evidence. Overall, the optimal model is the one that best aligns with the set of conjugation events as a whole. In some cases, it is unclear how many events happened. In particular, if a transconjugant divides before RFP fluorescence reaches threshold, each daughter lineage will correspond to a separate event. For the purposes of model comparison, we include one query for each transconjugant that reaches the fluorescence threshold. 

\subsection{Query Structure}

We query each model for the probability that transconjugant cells reach the RFP fluorescence threshold at the observed times. Each query is conditioned on the experimental data for all nodes outside the lineage of the transconjugant. That is, we provide the state of the entire model except for information from the lineage of the cell which was conjugated to as evidence. (Providing future information about the lineage would imply that a conjugation event had occurred.)

The selected expression delay distribution in each model gives rise to a range of possible times where conjugation could have occurred. For the sake of efficient computation, we omit the RFP nodes from our BNs in our implementation, and instead include them implicitly. Thus, a query is the combined probability of the transconjugant receiving the plasmid at any point in the range implied by the time it reached the RFP threshold, modified by the likelihood that the expression delay was exactly the implied length. (The range is based on the expression delay function.) Notably, we do not need to determine exactly when the conjugation event happened or which cell acted as the donor. This information is not useful for model comparison because there is no way to validate it against the experimental data.

Importantly, this query structure does not favor models that predict conjugation events early in the experiment because a lineage can only reach the RFP threshold once. Likewise, it does not favor models that assume every potential conjugation succeeds because the conjugation edge weights are normalized. It could, however, bias towards models with shorter maturation times. Earlier maturation produces more potential donors, which results in more opportunities for conjugation to occur. To compensate for this bias, we apply a second normalization based on a naive approximation of the likelihood of being mature at a given time. The approximation uses information about RFP expression to estimate when a conjugation event occurred. This normalization scheme results in a minor bias toward models that have longer maturation delays, but such models will still perform more poorly if they miss important opportunities for conjugation. 

\subsection{Comparison Metrics} \label{sec:metrics}

A meaningful model comparison metric must consider how a model performs on the data as a whole. Each microfluidic trap can be thought of as a single experimental trial that includes a number of conjugation events. To compare models, we need a metric that evaluates how each model performs on each trial and across multiple trials. The probabilities our models assign to individual conjugation events vary by many orders of magnitude ($10^{-28}$ to $10^{-4}$). The important information is not the magnitude of the probability assigned by each model, but the relative difference between them. Traditional methods, such as sum of squared errors, are not suitable because they consider the magnitude of the values. We consider two metrics for model comparison that capture different aspects of how the models assign probability. An ideal model would perform well on both.

The first metric we consider assesses which model performs the best for the most events. Our Average Trial Ranking is based on a group tournament ranking system. Suppose there are $k$ models of each trial. For each trial, we rank the models on each query from $1$ to $k$ based on the probability assigned to it. We then compute the average ranking of each model across all the queries in the trial. Finally, we average those values across all the trials. More formally, the Average Trial Ranking of a model $m$ is given by

$$avg_T(m) = \frac{1}{n} \sum_{t=1}^{n} \left( \frac{1}{q_t} \sum_{i=1}^{q_t} rank_i(m) \right),$$
where $n$ is the number of trials, $q_t$ is the number of queries in trial $t$, and $rank_i(m)$ is the rank of model $m$ on query $i$. A lower number corresponds to a better ranking. 

Because the probabilities are normalized, this approach may favor models that miss many events. Such a model could assign higher probabilities to the remaining ones. A model that consistently assigns probability to more observed events is arguably better.
The Total Probability Ranking involves a single averaging step. For each trial, we rank models from $1$ to $k$ based on the total probability assigned to all the queries in that trial. We then compute the average ranking of each model across all the trials. More formally, the Total Probability Ranking of a model $m$ is given by 

$$avg_P(m) = \frac{1}{n} \sum_{t=1}^{n} rank_t(m),$$

where $n$ is the number of trials and $rank_t(m)$ is the rank of model $m$ on trial $t$. A lower number again corresponds to a better ranking. 

\section{Results} \label{sec:results}

As this work is a proof-of-concept for our method, we focus on demonstrating the potential of our framework. We test two different versions for each type of function we can vary, producing a total of eight models. These are each tested on three experimental trials to assess consistency. The code for the model implementation is written in Python and is available on GitHub~\cite{ourgithub}. 

\subsection{Conditional Probability Distributions} \label{sec:functions}

The conjugation probability function can be used to investigate how the positioning of two cells impacts the likelihood of conjugation. Recall that a donor cell extends a pilus towards the recipient cell during conjugation. When considering the physical proximity of cells, we consider a `contact range' around the potential donor. Version one of the contact range function can be thought of as a baseline; it only checks if any part of the potential recipient is within that contact range. Because the pilus connects to the cell wall of the recipient, our second version measures how much of the edge of the recipient is within the contact range. Note that all contact calculations are done with approximate bounding boxes for computational efficiency. 

There is relatively little information about the length of the delays in the literature, so varying the length of our delay functions could provide valuable insight~\cite{massoudieh07}. Observations of similar systems in the literature suggest that maturation occurs within 40-90 minutes~\cite{andrup99,cullum78}, while~\cite{andrup98} suggests that most cells are mature by 40-45 minutes. To investigate, we consider two different ranges: 30 to 90 minutes and 15 to 75 minutes. (We chose a smaller lower bound for the first range because our model enforces a strict cutoff.) The ranges for the expression delay are based on manual observations of our data, from which we concluded that it was likely between 30 and 150 minutes. While long delays were possible, there was no clear evidence of them. We tested a second range of 30 to 120 minutes to examine whether long expression delays are likely. For simplicity, we use uniform probability distributions for all the delays. (Our framework works for any choice of probability distribution.) 

\subsection{Ranking}

We rank our models according to the two metrics described in Section~\ref{sec:metrics}. The results of the Average Trial ranking are shown in Table~\ref{tab:avg_trial}; the results of the Average Query Ranking are shown in Table~\ref{tab:total_prob}. Table~\ref{tab:compare_rank} compares how the different metrics ranked the models. These tables exclude queries that were incalculable due to memory constraints on the computers used. Furthermore, we exclude any query that was determined to be impossible by all tests. These may be due to tracking errors in the data, incorrect detection of transconjugants, or conditional probability distributions that eliminate actual conjugation events. More investigation is required to evaluate such cases. 

We manually checked the imaging data for one trial and determined that almost all impossible queries corresponded to tracking errors and/or incorrectly labelled transconjugants. Thus, our model identified multiple queries corresponding to conjugation events that were artifacts of errors in the data. The remaining queries all corresponded to a single lineage, which did appear to be near a donor cell. It is possible that our contact range was too small, or that the image processing steps did not accurately capture the edge of the cells. 

We use the convention Contact\_Expression\_Maturation to label each model. We refer to the functions described above in Section~\ref{sec:functions} as follows:
\begin{itemize}
    \item Base measures whether the recipient is in contact range of the donor.
    \item Edge measures how much of the recipient's edge is in contact range of the donor.
    \item R($l$,$u$) is a uniform distribution with lower bound $l$ and upper bound $u$ and represents the potential range of the expression delay. 
    \item M($l$,$u$) is a uniform distribution with lower bound $l$ and upper bound $u$ and represents the potential range of the maturation delay. 
\end{itemize}

The Average Trial Ranking suggests that the most significant factor for model accuracy is the conjugation probability function. Interestingly, this differs from the Total Probability Ranking which suggests that the most important factor is the expression delay function. Both models agree that the best conjugation probability function is the baseline and the best expression delay function is $R(30,150)$. According to the Average Trial Ranking, the maturation delay function is $M(15,75)$ is slightly better. The Total Probability Ranking does not distinguish clearly between the two functions. Both metrics rank models consistently across the three trials. 

\begin{table}[]
\centering
\begin{tabular}{l|cccc}
                          & \multicolumn{3}{c}{Trial} &         \\
Model                     & 1       & 2      & 3      & Average \\ \hline
Base\_R(30,150)\_M(15,75) & 3.65    & 2.36   & 3.4    & 3.13    \\
Base\_R(30,120)\_M(15,75) & 3.92    & 3.25   & 3.25   & 3.47    \\
Base\_R(30,150)\_M(30,90) & 3.86    & 3.69   & 4.19   & 3.91    \\
Edge\_R(30,150)\_M(15,75) & 4.67    & 4.17   & 4.66   & 4.50    \\
Base\_R(30,120)\_M(30,90) & 4.26    & 4.97   & 4.70   & 4.64    \\
Edge\_R(30,120)\_M(15,75) & 4.83    & 5.22   & 4.48   & 4.84    \\
Edge\_R(30,150)\_M(30,90) & 5.00    & 5.68   & 5.16   & 5.28    \\
Edge\_R(30,120)\_M(30,90) & 5.27    & 6.61   & 5.76   & 5.88   
\end{tabular}
\caption{\label{tab:avg_trial} The average ranking of each model across all the queries in a each trial. The last column is the Average Trial Ranking. All values are rounded to two decimal places.}
\end{table}

\begin{table}[]
\centering
\begin{tabular}{l|cccc}
                          & \multicolumn{3}{c}{Trial} &         \\
Model                     & 1       & 2      & 3      & Average \\ \hline
Base\_R(30,150)\_M(30,90) & 1       & 2      & 1      & 1.33    \\
Base\_R(30,150)\_M(15,75) & 2       & 1      & 2      & 1.67    \\
Edge\_R(30,150)\_M(30,90) & 3       & 4      & 5      & 4       \\
Base\_R(30,120)\_M(15,75) & 5       & 5      & 3      & 4.33    \\
Edge\_R(30,150)\_M(15,75) & 4       & 3      & 7      & 4.67    \\
Base\_R(30,120)\_M(30,90) & 6       & 6      & 4      & 5.3     \\
Edge\_R(30,120)\_M(15,75) & 7       & 7      & 6      & 6.67    \\
Edge\_R(30,120)\_M(30,90) & 8       & 8      & 8      & 8      
\end{tabular}
\caption{\label{tab:total_prob} The ranking of each model for each trial using the total probability assigned to queries. The last column is the Total Probability Ranking. All values are rounded to two decimal places}
\end{table}

\begin{table}[]
\centering
\begin{tabular}{l|cc}
                           & \multicolumn{2}{c}{Ranking}       \\
Model                      & Average Trial & Total Probability \\ \hline
Base\_R(30,150)\_M(15,75)  & 1             & 2                 \\
Base\_R(30,120)\_M(15,75)  & 2             & 4                 \\
Base\_R(30,150)\_M(30,90)  & 3             & 1                 \\
Bound\_R(30,150)\_M(15,75) & 4             & 5                 \\
Bound\_R(30,120)\_M(15,75) & 5             & 7                 \\
Bound\_R(30,150)\_M(30,90) & 6             & 3                 \\
Base\_R(30,120)\_M(30,90)  & 7             & 6                 \\
Bound\_R(30,120)\_M(30,90) & 8             & 8                
\end{tabular}
\caption{\label{tab:compare_rank} A comparison of the order in which each metric ranked the models.}
\end{table}

\section{Discussion} \label{sec:discussion}

The rankings disagree on the relative importance of the mechanisms they model. This discrepancy is likely due to what the metrics measure. Because the delay functions are uniform distributions, the most relevant factor for the probability assigned to an individual event is the probability of conjugation. Individual queries with higher probabilities will therefore correspond to the contact edges with the highest weights, and be ranked higher in the Average Trial Ranking. Conversely, the Total Probability Ranking should be less sensitive to individual events and more sensitive to the set of all possible events. It is therefore reasonable that a function that determines the possible events is most impactful for it.

Although the relative importance of the functions differs between models, they generally agree on which functions best represent the data. The lack of distinction between the two maturation delay functions may suggest that cells are maturing throughout both ranges. Overall, the results suggest that the delay before reaching RFP threshold may be much longer than the delay before maturation. This conclusion is further supported by the fact that a small number of queries ($\sim5$\%) returned zero only for models with the shorter expression delay and/or longer maturation delay. That is, some conjugation events could only be explained with a longer expression delay and/or shorter maturation delay. 

A more surprising result is the ranking of the conjugation probability functions. There is experimental evidence that the type of contact between cells impacts the probability of conjugation \cite{couturier23,goldlust23,lawley02,seoane11}. We expected the function accounting for the edge length in contact range to be more accurate than the baseline function. One explanation is that donors `select' recipients based on other features. One study investigated conjugation of F-like plasmids to multiple recipient species \cite{frankel23}. They found that donors select recipients based on proteins expressed on the cell surface. While our experimental system is different, we may be observing a similar phenomenon. (Recipient cells may have different concentrations of surface proteins.) Alternatively, studies suggest factors including growth rate, the stage of the cell cycle, and the relative orientations of cells affect conjugation frequency \cite{couturier23,goldlust23,lawley02,seoane11}. If these factors are more impactful than the length of the recipient edge near the donor, it would explain why the baseline conjugation function was more accurate.

\subsection{Future Work}

One significant challenge when working with imaging data is errors in image processing. These errors affect both the accuracy of our results and our ability to compute queries efficiently.
Moreover, our practice of passing transconjugant status down the lineages compounds errors from lineage tracking. For example, if a transconjugant cell is assigned a recipient as a child, the model will assume a recipient cell can act as a donor.

Additionally, it would be valuable to test our method on a larger set of trials to confirm that it is consistent and on a larger variety of models to gain more insight into conjugation. We could consider other biologically motivated conjugation probability functions that account for cell orientation and growth rate. It would be particularly interesting to identify a function that performs better than the baseline probability function. One could also attempt to find the most accurate delay ranges or consider different probability distributions within the ranges. Moreover, examining the applicability of this approach to other bacterial species and/or plasmid systems could further verify our results and provide insight about what mechanisms may differ between species and plasmid systems. 

The modelling approach taken in this paper, including our method of efficiently querying PGMs, may be applicable to other domains. It is often challenging to evaluate a PGM due to the complexity of the underlying calculations. The binary Noisy-OR structure described here allows for the efficient computation of a PGM over hundreds of thousands of nodes. Our approach may prove particularly useful for other populations in which the interactions depend on spatio-temporal factors. Other applications of the concepts and systems proposed in this paper include generalizations to similar networks in health or social sciences. The setup designed for horizontal gene transfer can be directly applied to modelling of the spread of a virus through a population such as a school. In this setting, contact metrics might measure the number of classes two people share or whether two people belong to the same friend group. A similar approach could be used to model the spread of ideas through a social network, again using social connections to determine the likelihood of transfer. Considering such applications could further the practical use of PGMs.

\section*{Acknowledgments}
This work was supported by funding from the Natural Sciences and Engineering Research Council of Canada (NSERC).  We thank Atiyeh Ahmadi for assistance with image processing.  This work was carried out
on the Haldimand Tract, land granted to the Haudenosaunee (Six Nations of the Grand River) in 1784.

\bibliography{library.bib}

\bibliographystyle{abbrv}

\end{document}